\documentclass[letterpaper,twocolumn,10pt]{article}
\usepackage{usenix}

\usepackage[dvips]{graphicx}

\usepackage{url,multirow}

\usepackage[dvipdfm,CJKbookmarks=true,bookmarksopen=true,colorlinks,citecolor=black,linkcolor=blue,anchorcolor=black,urlcolor=black] {hyperref}
\usepackage{balance}

\begin{document}

\date{}


\title{\LARGE \bf ACI{\huge \emph{A}}, not ACI\emph{D}: Conditions, Properties and Challenges}

\author{
{\rm Yuqing Zhu$^\star$, Jianxun Liu$^\star$, Mengying Guo$^\star$, Wenlong Ma$^\star$, Guolei Yi$^\dag$, Yungang Bao$^\star$}\\%
$^\star$Institute of Computing Technology, Chinese Academy of Sciences\quad\quad\quad$^\dag$Baidu
} 

\maketitle

\thispagestyle{empty}

\subsection*{Abstract}
Although ACID is the previous golden rule for transaction support, durability is now not a basic requirement for data storage. Rather, high availability is becoming the first-class property required by online applications. We show that high availability of data is almost surely a stronger property than durability. We thus propose ACIA (Atomicity, Consistency, Isolation, Availability) as the new standard for transaction support. Essentially, the shift from ACID to ACIA is due to the change of assumed conditions for data management. Four major condition changes exist. With ACIA transactions, more diverse requirements can be flexibly supported for applications through the specification of consistency levels, isolation levels and fault tolerance levels. Clarifying the ACIA properties enables the exploitation of techniques used for ACID transactions, as well as bringing about new challenges for research.\vspace{-6pt}

\section{Introduction}\vspace{-6pt}
Transaction support has been recognized again as indispensable for online applications in recent years~\cite{nosqlStonebraker}. Not implementing transactions in highly available datastores is even considered one's biggest mistake~\cite{deanMistake}. In recent years, systems like Megastore~\cite{megastore} and Spanner~\cite{spanner} emerge; and, academic solutions for transactional support are also proposed, e.g., MDCC~\cite{mdcc}, replicated commit~\cite{rcommit}, and TAPIR~\cite{tapir}. These emergent proposals guarantee the ACID properties of transactions in distributed replicated datastores. More importantly, they simultaneously consider the guarantee of high availability through data replication.

High availability is now the de facto first-class property required by online applications. Without high availability, even the temporary inaccessibility of data or service can lead to great economic loss~\cite{msOutEurope,serviceDown,amazonDown}. High availability is once guaranteed by abandoning ACID (Atomicity, Consistency, Isolation and Durability) transactions and supporting only BASE (Basically Availability, Soft state and Eventual consistency) data access~\cite{baseDan}. As the CAP Theorem~\cite{cap} states that only two can be guaranteed among consistency, availability and partition tolerance, the developers of BASE systems trade consistency for availability, guaranteeing eventual consistency instead of strong consistency and transactions~\cite{amazon07,eventual,dynamo}. In fact, the CAP Theorem does not indicate that transactions must be relinquished for high availability, which is then clarified~\cite{abadicap,brewercap2,raghucap,consistencydev}. Efforts are thus devoted to supporting transactions with high availability in recent years~\cite{granola,chains,paxoscp,msgfutures,helios,tapir,rcommit,spanner,megastore}. A concept of highly-available transactions is even proposed~\cite{hat,hatvldb}. Transactions now must be supported with high availability for online applications.

In this paper, we propose ACIA to be the new standard for transaction support, instead of ACID, replacing \emph{Durability} with \emph{Availability}. As demonstrated by the years of practice with the BASE model, applications can work well with systems and data in soft state~\cite{salt,baseDan,hbase,dynamo,cassandraPaper}, even on the datacenter-scale power outage~\cite{awsoutage}. Soft state only requires that correct data or states can be regenerated at the expense of additional computation or file I/O on faults such as network partition~\cite{baseBrewer}. With soft state guaranteeing availability, durability is no longer a fundamental property required by data management. In fact, as long as data is available, applications do not care about whether the correct data is durably stored or regenerated on the fly in the system.

We show that high availability of data is in fact a stronger property than durability (\emph{Section~\ref{sec:proof}}). Highly available data can be made durable, while durable data is not necessarily highly available. Hence, ACIA transactions cannot be supported by all proposals designed for guaranteeing ACID transactions over replicated datastores. The fundamental reason is that ACIA transactions assume new conditions commonly made for online applications in system implementations (\emph{Section~\ref{sec:conditions}}). These conditions are different from those generally assumed for the classic data management systems~\cite{concurrencybook}. Assuming the classic conditions, e.g., predictable communication delay, some recent proposals~\cite{helios} cannot support ACIA transactions. Even some proposals can support ACIA transactions, they are not necessarily designed without redundant components, e.g., persistent logs~\cite{mdcc,tapir}. We present a specification of ACIA properties, which constitute the major highlights of a new transaction paradigm (\emph{Section~\ref{sec:properties}}). To check whether the transaction is supported by a particular system, one only needs to make an ACIA test of the system's quality. Clarifying the ACIA properties enables implementations to combine different mechanisms that guarantee the properties respectively. Besides, it also enables us to explore the new space of research challenges and problems (\emph{Section~\ref{sec:challenges}}).\vspace{-6pt}

\section{High Availability vs. Durability}\vspace{-6pt}
\label{sec:proof}
High availability is a property specifying that a database, service or system is continuously accessible to users~\cite{baseBrewer}. Using with transactions, high availability refers to a property of database in this paper. We say that a database is high available if any client connected to the system can access any data within the database at any time. Note that, this description does not concern about implementations. In comparison, the durability property of ACID requires that a committed transaction's effect be reflected by the database in the persistent storage, such that the effect of the committed transaction will not be lost even on power failures~\cite{acid}.

As a property of the database, high availability shares a few similarities with durability. First, high availability is a property independent of the atomicity, consistency, and isolation properties. Each of the four properties must be guaranteed by the respective measures. Second, high availability places no constraints on data or replica consistency~\cite{graytxnbook,consistencyLevels}. A client accessing the highly available database can find the data outdated or up-to-date; and, how inconsistent the returned data can be depends on the consistency level agreed between the client and the database. Third, high availability has no indications for isolation levels. ACID transactions can be implemented with different isolation levels~\cite{critique}, while different isolation levels can result in different database states kept durable. Similarly for highly available database, the isolation levels supported by the system can also affect the database image observed by a client.

High availability of data is almost surely a stronger property than durability. On the one hand, highly available databases can be made durable easily. Given high availability of data, we can use a client program to traverse the whole database and store all returned values to a persistent storage, leading to durability. Nowadays, highly available systems usually distribute their storage across multiple geographic locations. It is highly unlikely that power failures will affect storage in all locations. That is, power failures will almost surely not affect highly available systems, unless the power failure is global (a case with probability zero).

\begin{figure}[!h]
\vspace{-6pt}
      \centering
      \includegraphics[width=0.3\textwidth]{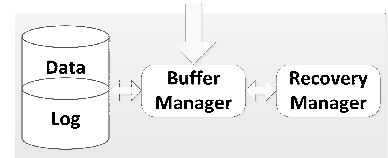}\vspace{-3pt}
      \caption{The typical architecture of a durable storage engine.}\vspace{-9pt}
      \label{fig:durability} 
\end{figure}
On the other hand, durable systems are not necessarily accessible at any time. To guarantee data durability, log-based crash recovery is extensively studied for ACID transactional database research~\cite{recovery,aries}. The typical architecture of such durable and recoverable database systems is demonstrated in Figure~\ref{fig:durability}. The four basic building blocks for such systems include durable data image, durable logs, buffer manager and recovery manager. The recovery manager relies on the durable logs to recover a correct database on errors, while the buffer manager manages durable data and logs for program access. The underlying assumption for such an architecture is that the recovery allows a system down time. In comparison, highly available systems handles failures or errors without system down time.

\begin{figure}[!h]
\vspace{-6pt}
      \centering
      \includegraphics[width=0.3\textwidth]{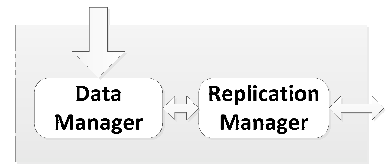}\vspace{-3pt}
      \caption{The typical architecture of a highly-available storage engine.}\vspace{-9pt}
      \label{fig:availability} 
\end{figure}
With high availability, a database will not become inaccessible because of partial system failures such as node failure or network partition. Replication is the typical mechanism to guarantee high availability of data. Exploiting fault-tolerant replication algorithms, partial system failures can be tolerated without bringing the system down. The number of replica failures a system can tolerate depends on the algorithms used and the consistency level specified. For example, eventually consistent systems can guarantee high availability as long as all replicas do not fail simultaneously~\cite{eventual}. A fault-tolerant system can be made highly available by fixing failed components timely. The architecture of a typical storage engine guaranteeing high availability is demonstrated in Figure~\ref{fig:availability}. The data manager allows the data to reside in memory or in persistent storage.\vspace{-12pt}

\section{New Implementation Conditions}\vspace{-6pt}
\label{sec:conditions}
Conditions previously assumed for distributed system implementations become invalid now for large-scale distributed systems that are widely deployed to support the plethora of online applications. One example is whether a system can be inaccessible for some time. Previously, systems can have a down time for failure recovery (MTTR, Mean Time To Recovery); now, the system must remain accessible for the high availability requirement. Another example is whether a system can have synchronized clocks across servers. Previously, system servers are assumed to have synchronized clocks, enabling the timestamp-based distributed concurrency control (CC)~\cite{dtcc}; now, the clocks on different clocks can be coordinated at most to a certain precision, requiring a different CC design~\cite{spanner}.

We capture the changed conditions for system implementations by observing how Consensus algorithms are exploited in replicated systems to guarantee high availability~\cite{chubby,paxosrsm,paxosrep2,livepaxos}. The most widely used Consensus algorithm is Paxos~\cite{paxos,livepaxos}. Years of practice have proven that Paxos is feasible in the practical scenarios of large-scale distributed systems. The Paxos algorithm assumes the asynchronous system model~\cite{systemModel} and crash-stop failure~\cite{failureModes}. The crash-stop failure is related to the conditions of \emph{implementation compliance} and \emph{node recovery moment}, while the asynchronous system model is related to the phenomena of \emph{unpredictable message delays}, \emph{inconsistent clocks} and \emph{unreliable failure detection}. The conditions and the phenomena must all be considered in system implementations.

\textbf{Implementation Compliance}. A server node in the system must behave as the implementation dictates. Besides, it must follow the protocol implemented by the system to send and receive messages. A node can either behave according to the implementation or crash to stay in a stop state. The above phenomenon happens when the crash-stop failure is assumed. Other failure modes also exist. A commonly studied failure is when the server can send arbitrary messages to other servers, without complying to the implementation. This failure mode is called the Byzantine failure~\cite{byzantine}. At present, the most commonly assumed failure is the crash-stop failure.

\textbf{Node Recovery Moment}. In large-scale distributed systems, the failures of individual server node are considered common situations, which must be tolerated by the system software. Replication is the common technique for fault tolerance and high availability. While the model of replicated state machines (RSM)~\cite{rsm} offers a measure to describe and analyze replication, Consensus algorithms coordinate the RSM to process commands properly even under node failures. The decision of each command for the RSM is modeled as a Consensus problem. Designed to solve the Consensus problem~\cite{consensus}, one run of a Consensus algorithm can reach a single agreement on the command for the RSM (denoted as \emph{single-decree} in a related work~\cite{raft}). Therefore, multiple runs of the algorithm are needed to make the system functional.

A node cannot recover and rejoin the system during a run of the Consensus algorithm. During a run, Consensus algorithms generally assume the static membership of nodes (called participants, acceptors or learners)~\cite{paxosSimple}. Nodes can leave the membership due to failures but not join the membership. Rather, the recovery and the change of membership must be handled \emph{before} or \emph{after} a run of the Consensus algorithm. Otherwise, extra mechanisms called reconfiguration~\cite{paxosreconfigure1,paxosreconfigure2} must be added to the implementation. In fact, a recovered node does not need to join the system during an algorithm run unless the number of failed nodes exceed the algorithm's fault tolerance level.

\textbf{Unpredictable Message Delay}. The delay for a message to reach its receiver is not predictable. This condition exists because large-scale systems supporting online applications can be globally distributed and the system network in wide area is highly unpredictable. It is possible that a message sent by one server travels in the system network for an indefinitely long time, making the message look like a lost message. Note that this condition also models the situation that messages can get lost for some undetermined reasons, e.g., network congestion. Implementations relying on predictable communication delays~\cite{helios} are not suitable for the real applications with this condition.

\textbf{Inconsistent Clocks}. Different servers in the same system can hardly have consistent clocks. The reason is two fold. First, the local clocks of servers can drift independently as time passes. Second, the communication delay is neither constant nor predictable. Although techniques exist to synchronize the clocks to a certain precision~\cite{spanner}, designs relying on precise timestamps~\cite{dtcc} will generally not work in highly available systems. To track the global ordering of events, mechanisms such as vector clocks are needed~\cite{rsm,dynamo,cops}.

\textbf{Unreliable Failure Detection}. A server cannot precisely distinguish message slowness from failures of other servers. Since a message can travel in the system for an indefinite long time, a server has no way to differentiate whether its message is slow or the interacting server has failed. Furthermore, if a server is not receiving messages from any other server, it can hardly tell whether all other servers fail, whether a network partition happens, or whether all messages are travelling slowly suddenly. None of the three conditions can be made certain of. That is, reliable failure detection is not possible, which then rules out all possible solutions~\cite{flp}. To solve the problem in practice, a server will resend its messages and take communication timeouts as the signal for server failures~\cite{bigtable,spanner,dynamo,cassandra,pnuts}. Therefore, the corresponding condition for failure detection is that server failures can be detected in some way such as timeouts.

\begin{table}
  \centering
  \small
  \caption{Implementation Conditions: ACID vs. ACIA.}\vspace{6pt}%
  \label{tbl:changes}%
  \renewcommand\arraystretch{1.2}
  \begin{tabular}{|l|c|c|}
\hline
   {\normalsize \textbf{Conditions}} & {\normalsize \textbf{ACID}} & {\normalsize \textbf{ACIA}}\\
   \hline\hline
   \textbf{System} & \multirow{2}*{With down time}  & \multirow{2}*{Without down time}\\
   \textbf{recovery} & & \\
   \hline
   \textbf{Message} & \multirow{2}*{Predictable}& \multirow{2}*{Unpredictable}\\
   \textbf{delay} & & \\
   \hline
   \textbf{Synchronized} & \multirow{2}*{Possible} & \multirow{2}*{Impossible}\\
   \textbf{clocks} & &\\
   \hline
  \textbf{On failure}  & \multirow{2}*{Stop and restart}& \multirow{2}*{Fault tolerance}\\
   \textbf{detection}&  &\\
\hline
\end{tabular}\vspace{-12pt}
\end{table}
\textbf{Conditions for ACIA vs. ACID}. We summarize the changes of typical implementation conditions from ACID to ACIA in Table~\ref{tbl:changes}. Four main changes exist. First, classic ACID designs assume the system recovers with down time, while ACIA requiring high availability assumes no system down time. Second, classic ACID designs assume predictable message delays to enable timestamp-based mechanisms, while ACIA assumes unpredictable message delays. Third, classic ACID designs are for system with limited distribution and synchronized server clocks, while ACIA are for large-scale systems that are impossible to have synchronized server clocks. Fourth, classic ACID systems stop and restart on failure detection and recovery, while ACIA systems assume built-in fault-tolerance mechanisms. These changes and differences are in general. There exist early systems assuming some of the conditions of ACIA systems, but not all conditions are considered. Now with ACIA requirements, all the conditions must be considered and handled in the system implementations.\vspace{-6pt}

\section{Specification of ACIA Properties}\vspace{-6pt}
\label{sec:properties}
Given the conditions of Section~\ref{sec:conditions}, the four properties of ACIA are specified as follows.\vspace{-6pt}
\begin{itemize}
  \item \textbf{Atomicity}. The effects of either all or none of the operations in a transaction are reflected in the database, with the user knowing which of the two results is.\vspace{-6pt}
  \item \textbf{Consistency}. Each successful transaction can commit only \emph{legal} results, preserving the consistency of the database. Legal results comply with rules and consistency levels specified for the database, e.g., integrity constraints~\cite{graytxnbook} and replica consistency levels~\cite{consistencyLevels}.\vspace{-6pt}
  \item \textbf{Isolation}. Operations within a transaction must be isolated from other transactions running concurrently. How much transactions can be isolated from other transactions is defined as the isolation levels~\cite{critique,adyaIso}, which are guaranteed by different concurrency control mechanisms~\cite{ingres,ccsurvey1,ccsurvey2}.\vspace{-6pt}
  \item \textbf{Availability}. Once a transaction has committed its results, the system must guarantee that these results are reflected in the database, whose data can be accessed by any client connected to the system.\vspace{-6pt}
\end{itemize}

\textbf{The Clients' View}. Previously with ACID, the client understands that transactions are atomic and durable and that the consistency and the isolation conditions can be specified for different applications. Typical consistency level agreements are integrity constraints like correlated updates for foreign keys~\cite{concurrencybook}. Typical isolation level agreements are like serializability, snapshot isolation or read committed~\cite{critique}.

With ACIA, the database is guaranteed to be continuously accessible and transactions are atomic. The various requirements of the client can be flexibly supported through the specification of consistency levels, isolation levels and fault tolerance levels. Different from ACID, the consistency levels for ACIA need to include replica consistency specifications. Besides, new isolation levels are possible and to be added~\cite{zhuIso}. The levels of fault tolerance can also be specified. The level of fault tolerance can be traded off with performance. For example, a low level of fault tolerance enables the system to use fast Paxos for low latency, while a high level can result in the system using the classic Paxos but with higher latency~\cite{paxos,fastpaxos}.

\textbf{Implementation}. Clarifying the ACIA properties enables us to exploit early research results. Similar to the implementation of ACID transactions, the support of ACIA transactions require the implementation of consistency compliance, concurrency control and commit coordination, as well as replica control. In the past, the former three aspects are discussed separately with different measures. For example, consistency compliance can be about how foreign keys can be updated efficiently; concurrency control can be about which isolation levels are best for an application and which scheme is most efficient for a workload; and, commit coordination is about how atomicity is guaranteed through different protocols like 2PC or 1PC. We can thus exploit and combine such mechanisms to guarantee the ACIA properties.

Although the four properties of ACIA must be guaranteed simultaneously for transactions, the guarantee of the first three properties are independent. As each of the three properties are guaranteed through different schemes, numerous combinations of the schemes are possible. Furthermore, the consistency and the isolation can be guaranteed with multiple choices, various combinations of these choices are possible, leading to flexible application usages. Current solutions to transaction support in highly-available systems usually interwind the mechanisms that guarantee these properties. For example, highly-available transactions~\cite{hat,feral} mix replica consistency and availability with isolation levels, leaving a vague image blurring replica control and concurrency control. Clarifying the properties of ACIA enables further exploration in system designs and implementations.\vspace{-12pt}

\section{Challenges}\vspace{-6pt}
\label{sec:challenges}

We now discuss the challenges in supporting ACIA transactions. These challenges arise mainly due to the changes of implementation conditions.

\textbf{Atomic Commit Problem Redefined}. The classic problem of atomic commit assumes each participant serves orthogonal data units for a distributed transaction~\cite{concurrencybook}. Many widely used protocols are designed to solve this classic atomic commit problem~\cite{aries,1pc,txnConsensus}. For ACIA transactions, multiple participants can server the same data units. These participants can fail independently, but not all participants serving the same data unit fail simultaneously. Obviously, the classic problem definition of atomic commit is not suit for ACIA transactions. A new problem definition of atomic commit is needed.

Furthermore, typical solutions to the classic atomic problem relies on persistent logs~\cite{spanner,aries,1pc}. With high availability of data, keeping durable logs is not a wise choice. Therefore, new mechanisms not using persistent logs need to be devised to solve  the new atomic commit problem of ACIA. In fact, as the classic atomic commit solutions is a vote-after-decide process~\cite{rstar}, we can take the vote-before-decide alternative and transform the new atomic commit problem into the Consensus problem~\cite{hacommit}, making a solution based on existent algorithms possible.

\textbf{Comprehensive Consistency Levels}. The consistency of ACID is enforced through the compliance check of all defined rules, including constraints, cascades, triggers, and any combination thereof. These rules are based on a single-replica database image. In comparison, ACIA databases are inherently distributed and replicated. Therefore, the rules of constraints, cascades, triggers, and their combinations must be extended to consider replicas. Furthermore, there exist multiple consistency levels for replicated data, and these consistency levels are allowed by different applications~\cite{consistencyLevels,consistencyrationing,fbConsistency1,fbConsistency2}. The following questions thus remain to be answered: (1) how can these replica consistency levels be combined with classic consistency rules to offer comprehensive consistency levels? (2) which of the comprehensive consistency levels are allowed by applications? and, (3) how can the new consistency levels be specified in an application-understandable manner?

\textbf{Application-Understandable Isolation Levels}. Classic concurrency control mechanisms study how operations of different transactions can be ordered correctly. For ACIA transactions, concurrency control must also consider how each operations can ordered on different replicas, leading to a more complex global view of operation executions. The phenomena-based specification~\cite{critique,feral,adyaIso} of classic isolation levels cannot cover all isolation levels possible for ACIA transactions, e.g., partition-based isolation levels~\cite{zhuIso}. The spectra of isolation levels possible for ACIA transactions need to be studied, and a corresponding specification is in need. Besides, the specification must be made application-understandable, as isolation levels are mainly for applications to choose the correct implementations.

\textbf{Failure Detection Mechanisms \& Fault-Tolerance Levels}. Failure detection is the basis for fault tolerance. Systems supporting ACIA transactions must have built-in fault tolerance mechanisms and thus failure detection mechanisms. Failure detections with different reliability require different solutions~\cite{failureDetector,failureDetectors,weakft}, which can tolerate varied numbers of failures (denoted as fault-tolerance levels). Current implementations for failure detections are mainly based on timeouts~\cite{falcon}. Systems implementing failure detectors with different reliability remain to be devised. And, a specification on what fault-tolerance levels supported by such systems needs to be provided.\vspace{-12pt}

\section{Conclusion}\vspace{-6pt}

This paper makes a key observation that the high availability requirement of data has changed the conditions for transaction support. In this paper, we have shown that high availability is almost surely a stronger property than durability. On proposing ACIA instead of ACID, we have investigated the conditions for system implementations supporting ACIA transactions. The change of implementation conditions leads to new challenges for transaction support in large-scale distributed systems. We analyze the challenges regarding each property of ACIA. The implementation-independent specification of ACIA properties not only enables the reuse of mechanisms previously devised to support ACID transactions, but also opens up a new research space for future exploration.

\textbf{Acknowledgments} This work is supported in part by the State Key Development Program for Basic Research of China (Grant No. 2014CB340402) and the National Natural Science Foundation of China (Grant No. 61303054).

\balance

{\footnotesize \bibliographystyle{acm}
\bibliography{ref}}


\end{document}